\documentclass[twocolumn]{article}

\usepackage{bm}
\usepackage[margin=.75in]{geometry}
\usepackage{graphicx}
\usepackage{epstopdf}
\usepackage{caption}
\usepackage{placeins}
\usepackage{titlesec}
\titleformat*{\section}{\large\bfseries}
\usepackage{algorithm}
\usepackage{algorithmic}
\usepackage{cite}
\usepackage{ulem}
\usepackage{color}
\usepackage{array}
\usepackage{hyperref}
\hypersetup{
	colorlinks=true,
	linkcolor=blue,
	urlcolor=red,
}
\begin{document}

\title{The Hydra String Method: A Novel Means to Explore Potential Energy Surfaces and its Application to Granular Materials}

\author{Christopher Moakler         \and
        Katherine A. Newhall 
}

\author{Christopher Moakler\\
University of North Carolina at Chapel Hill \\
Department of Physics and Astronomy, Chapel Hill, NC         \\ and \\
        Katherine A. Newhall \\
        University of North Carolina at Chapel Hill\\ Department of Mathematics, Chapel Hill, NC
}

\date{\today}

	\maketitle

		\begin{abstract}
			We present a novel means to understand granular materials, the Hydra String Method (HSM). This is an efficient and autonomous way to trawl an arbitrary potential energy surface (or any similarly high dimensional function) that enumerates the saddle points, minima, and minimum energy paths between them. In doing so, it creates a reduced dimensional network representation of this surface.  We also present a series of tests to choose optimized parameters for the application of the HSM.  We apply this to the potential energy function of a granular system consisting of a configuration of bi-disperse, frictionless, soft spheres. Future work will make use of the found ensemble of transition pathways to statistically predict the dynamics of a system of grains. 
			
		\end{abstract}

\section{Introduction}	\label{Intro}
The class of granular materials is second only to water in materials used by humans and despite recent advancements to understand their unusual behavior, scientists still lack an adequate way to describe the full range of their behavior \cite{Handbook,Jaeger_1996,Richard_2005,van_Hecke_2009}.  Attempts at firmly grasping the dynamics and mechanics of granular materials have been as fruitful as firmly grasping a handful of sand, it quickly runs away from us.
	
This work presents the foundation of a new approach to predict the dynamic response of purely repulsive-interaction granular systems from a defined energy landscape.  We take a soft-sphere quadratic potential between overlapping 2D bidisperse frictionless disks and map out energy-minimizing jammed states as well as the paths between neighboring states that minimizes the energy, the minimum energy paths (MEP).  Unlike other work that seeks to enumerates all uniquely jammed states~\cite{Gao_2009,Ashwin_2012}, we only seek a representative sample of states as full enumeration is computationally infeasible for even small systems of 12 particles. Note that the neighboring states and pathways between them are dependent on the protocol generating the packing (c.f.~ref.\cite{Tuckman_2020} that allows the box size to change).  We hope this energy-based approach will lead to predicting the dynamic response of the system, like the mapping generated from observing rearrangements under strain as in ref.\cite{Mungan_2019}.  

\begin{figure}[t]
		\centering
		\includegraphics[width=0.95\linewidth]{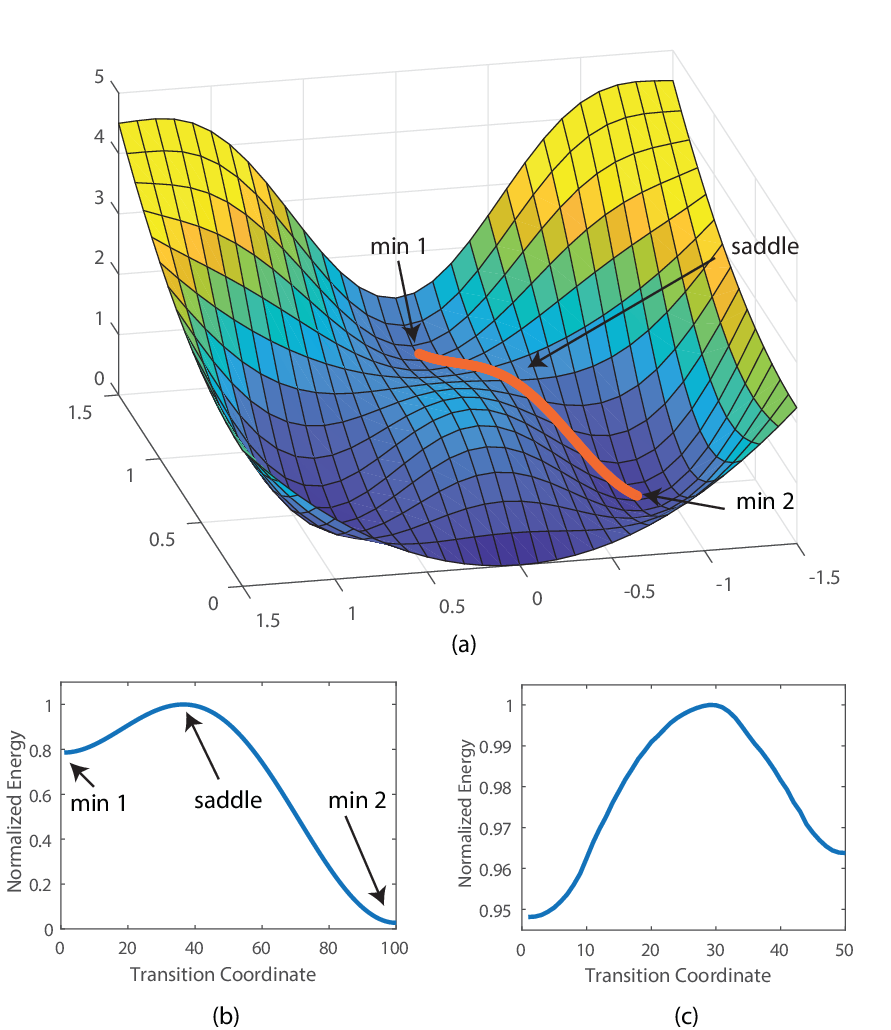}
		\caption{(a) A simple potential function with a MEP shown in orange. (b) The normalized energy along the transition path or ``coordinate" (c) Example of an energy vs. transition coordinate for a granular system moving between minima.}
		\label{fig:TransCoord}
	\end{figure}

The idea of exploring an energy landscape is not itself new.  In fact the building blocks of our automated ``Hydra'' method are the string method \cite{E_2002} and its climbing variant \cite{Ren_2013} (two of many such methods~\cite{Henkelman_2000,Henkelman_1999,Olsen_2004})
developed for use in computational chemistry to probe the evolution of a single reactant to a product state with, possibly many, intermediate transition states.  These transitions are often diagrammatically sketched as a graph of energy vs. reaction coordinate and called reaction coordinate diagrams~\cite{M_ller_1980}.  Intermediate transition states along this path appear as maximums in energy, but correspond to saddle points in the full high-dimensional phase space of the system.  Such a saddle point sits on top of the local lowest energy barrier separating two states and therefore chemical reactions are most likely to proceed over this barrier near a MEP (c.f.~\cite{Vanden_Eijnden}).  Although granular systems are not perturbed by thermal noise, one might expect a slowly-sheared granular system to follow pathways near these low energy saddle points, possibly following an MEP like the one shown in Figure \ref{fig:TransCoord} for a simple 2D potential (a and b) and for a more complicated soft-sphere model (c) discussed in more detail in Sec.~\ref{Test System}.

Our automated method reduces the energy landscape to a network representation of transition pathways between neighboring states, through saddle points, like those presented for smaller systems such as six colloidal particles with depletion attraction \cite{Perry_2015}, nematic liquid crystals on 2-D hexagons \cite{Han_2020}, small atomic clusters interacting with Lennard-Jones potential \cite{Doye_2005,Doye_2002} and 2D lattice polymers  \cite{Sch_n_2002}. Unlike previous approaches, we find not just the transition state but the full transition path. The method sends out multiple climbing strings to find saddle points that are ensured to be on the edge of the basin of attraction around a minimum by the monotonicity of the energy along the string.  From the unique saddle points found, strings are descended to find neighboring minima.  This process is repeated and, like a Hydra, this ``string'' will grow new exploring heads to continue the energy landscape mapping.

The main focus of this paper is laying out a sequence of tests that can be applied to any system with an energy landscape to determine parameters that balance accuracy with computational cost to be able to efficiently sample higher-dimensional more complex systems.  These tests seek general characteristics of the basins of attractions around each energy-minimizing state in order to efficiently climb out of the basins and sample the neighboring states.  Knowing the distance between basins dictates how precisely each state must be resolved to determine if two states are unique.  Determining the characteristic radial size of a basin informs how to resolve the smallest features of the landscape and how far away to look for saddle points on the edge of this basin.  The last test determines how many saddle points should be sampled to have a reasonable sample of unique saddle points on the edge of the basin of attraction.

The remainder of the paper is organized as follows.  
In Sec.~\ref{sec:Hydra} we briefly review the string and climbing string methods and present the details of our Hydra string method.
We explain the three main tests to apply to a system in order to determine parameters for the Hydra string method that will balance accuracy with computational efficiency in Sec.~\ref{sec:NumSims}.  
In Sec.~\ref{sec:Results} we apply these tests to the granular system given in 
Sec.~\ref{Test System}.
We conclude the paper with discussions of various parameter choices in Sec.~\ref{sec:Discussion} and conclusions in Sec.~\ref{sec:Conclusion}.


	\section{The Hydra String Method and Algorithm\label{sec:Hydra}}
	
	The Hydra String Method (HSM) is an extension of the string and climbing string methods designed to explore the potential energy surface (PES) in a systematic way, creating a network of connected minima and intermediate saddle points.  We briefly review the string and climbing string method before proceeding to describe our algorithm in more detail.
	
		\subsection{The String Method}

	The String Method finds the MEP between two minima but is more computationally efficient and exhibits greater stability than the popular Nudged Elastic Band (NEB) method \cite{Mills_1995}.  Both methods evolves copies, or ``images", of the system in configuration space by gradient descent.  To address the issue of the images falling towards the minima on either end of the path and bunching up there, the NEB method introduced springs between the images to keep them separated along the MEP. These spring forces couple the images that would otherwise independently undergo gradient descent. The String Method decouples these competing forces on the images and treats them as two different steps, thus enhancing efficiency. The String Method first allows the images to follow a gradient descent for one time step and then interpolates the images to equally space them along the string in arclength.
	
	One of the benefits of decoupling these two processes is that it allows the user to change the interpolation method ``on the fly". A linear interpolation method can easily be changed to, for example, a natural spline method or any other desired interpolation method. The decoupled images also do not effect one another before the interpolation, so they can be computed in parallel. 
	
	The string method can be implemented with either one or both ends of the string held fixed. When both ends are held fixed the string methods converges to a MEP between those two points while keeping only one end point fixed allows the string method to converge to a MEP between a known minimum and a newly found minimum.
	This method was designed as a tool for computational chemistry where one may know both end points (known reactant and product states) or with only a single known end point (known reactant or product state).
	
	\subsection{The Climbing String Method}

	The Climbing String Method \cite{Ren_2013} is a modification of the String Method that finds saddle points from a single known minimum by allowing the final image along the string to climb up against the gradient with the following force acting on it:
		
	\begin{equation}
		\label{Climbing String Force}
		\textbf{F}_{i_{cli}} = -\boldsymbol{\nabla} V(\textbf{r}_{i_{cli}}) + \nu(\boldsymbol{\nabla} V(\textbf{r}_{i_{cli}}),\hat{\tau}_{i_{cli}}) 	\hat{\tau}_{i_{cli}}.
	\end{equation}
	Here, $\boldsymbol{\nabla} V$ is the gradient of the potential energy, $\textbf{r}_{i_{cli}}$ is the configuration of the system at the climbing image, $\nu$ is a tunable parameter (generally set to 2) that controls the climbing speed of the string, and $\hat{\tau}_{i_{cli}}$ is a unit tangent approximation to the string at the climbing image.

	The first image on the string is kept fixed at a minimum and the intermediary images follow a gradient descent at each time step. As in the String Method, the intermediary images are redistributed via an independent interpolation. A computationally cheap option for this redistribution is a simple linear interpolation. But, other methods can easily be substituted.

	To ensure that the string remains in the basin of the starting minimum, if, at any time step, the energy along the string fails to be monotonically increasing, the string is cut at the point of non-monotonicity. The images are then re-interpolated along this shorter string and the process is continued until it converges to a MEP.

	\subsection{The Hydra String Method}
	
\begin{figure*}
	\centering
	\caption{Depiction of the HSM on the 2D PES, $E = x^2 + y^2 + \sin(\pi x) + \cos(\pi y)$, shown as a contour plot. (a) Evolving climbing strings with one end pointed pinned at the original minimum. (b) Convergence of the strings to minimum energy paths ending at two new saddles. (c)	and (d) Descent from these saddles to new minima. (e) Evolving and (f) convergence of climbing strings from one of the newly found minima.}
	\label{SnakeingAlgPic}

	\includegraphics[width=\linewidth]{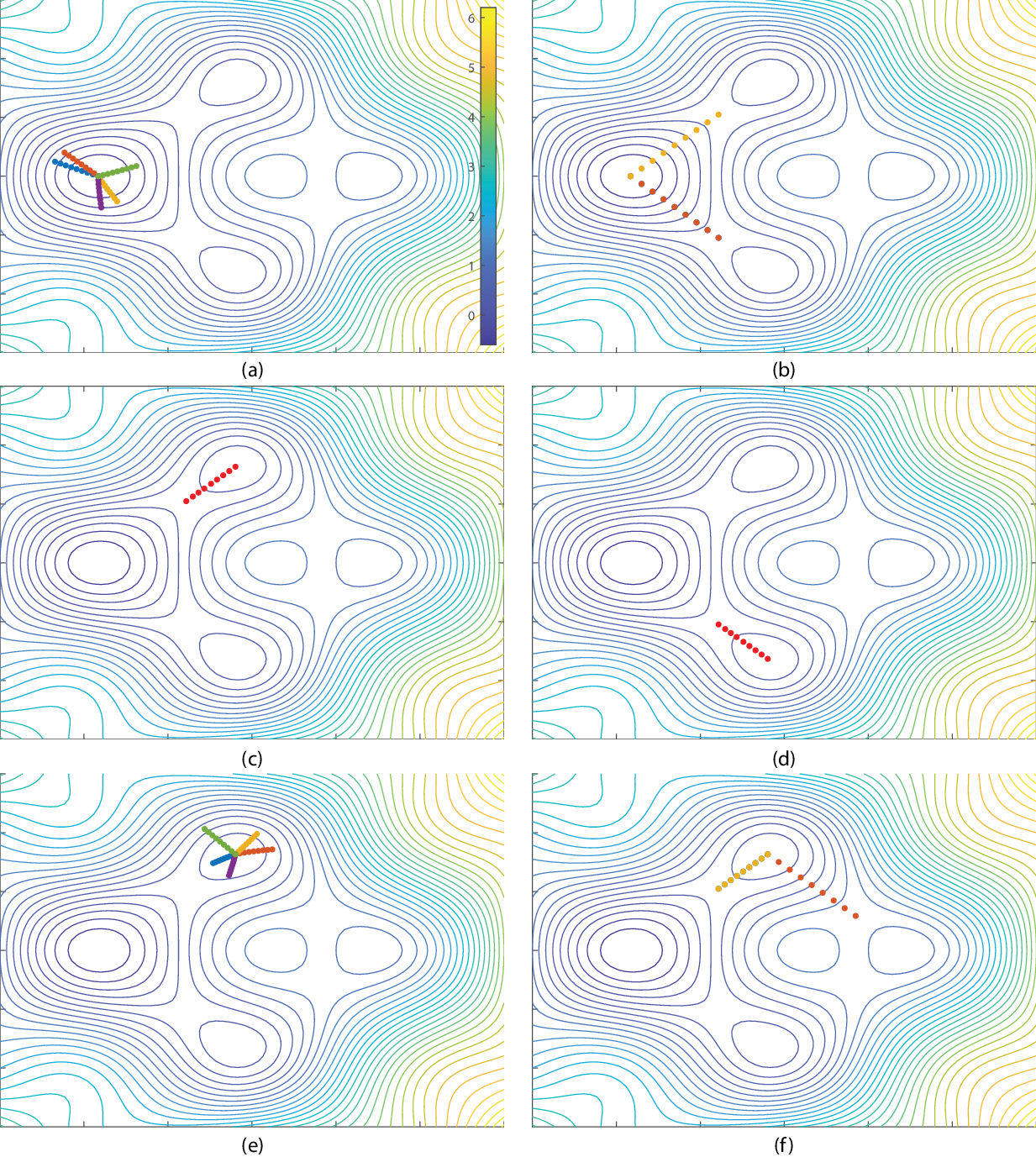}

\end{figure*}

	We describe the Hydra string method in more detail, which is illustrated in Fig.~\ref{SnakeingAlgPic} and summarized by the pseudo-code in Algorithm \ref{SSAlg}. Its \textsc{Matlab} implementation can be found on GitHub, see Code Availability below

	An initial minimum is found by choosing a random point and then performing a gradient descent. This minimum is added to a running list of minima found on the PES. From this initial minimum, a number of new strings, $n_{strings}$, with $n_{images}$ images along each string are extended a given distance, $dist_{ext}$, in a random direction and allowed to climb following the climbing string method described above. (A set of initial strings evolving are shown in Fig.~\ref{SnakeingAlgPic}.a). Those strings that converge to a saddle point (see the converged strings in Fig.~\ref{SnakeingAlgPic}.b) have their saddle points added to a running list of saddle points. The images on these saddle points are then perturbed, away from the minimum from which they were found, and allowed to follow a gradient descent until they converge to another minimum (see examples of such a converged path in Figs.~\ref{SnakeingAlgPic}.c and d). These new minimum are also added to the running list of minima. An unexplored minimum is chosen from the list and the process begins again (see new evolving strings in Fig.~\ref{SnakeingAlgPic}.e and converged strings in Fig.~\ref{SnakeingAlgPic}.f). This procedure is repeated until either no new minima/saddles are found or a predetermined number of minima are explored. 	In practice, one only wants to add unique minima(saddles) to the list of minima(saddles). Thus at every step where new minima(saddles) are found, they must be compared to the preexisting list of minima(saddles) with a tolerance, $tol_{sad/min}$, to determine uniqueness before being added to the list.

	\begin{algorithm}[]
		\caption{Hydra String Algorithm}
		\label{SSAlg}
		\begin{algorithmic}[1] 
			\STATE $x_{random}$ \%randomly chosen position
			\STATE $x_{initial}$ = Descend$(x_{random})$ \%descend to initial minimum in system
			\STATE $x_{min_{new}}$ = $x_{initial}$ \%add first minimum to list of unexplored minima
			\STATE $x_{min_{unique}}$ = $x_{initial}$ \%add first minimum to list of unique minima
			\item[]
			
			\WHILE{$x_{min_{new}}$ is not empty}        
			\STATE $x_{min}$ = $x_{min_{new}}(1)$ \%Pick next minimum to explore
			\STATE delete $x_{min_{new}}(1)$ \%Remove minimum to be explored from unexplored list
			\item[]
			\FOR {$i$ = $1$:$n_{strings}$}
			\STATE $x_{min_{extended}}$ = Extend$(x_{min}$,$dist_{ext})$ \%Extend a string in a random direction
			\STATE $possible_{unique\_sad}(i)$ = Climb$(x_{min_{extended}})$	
			\item[]
			
			\IF {$possible_{unique\_sad}(i)$ is unique}
			\STATE$x_{sad_{temp}}$.append$(possible_{unique\_sad}(i))$
			\ELSE
			\STATE continue
			
			\ENDIF
			\item[]
			\ENDFOR
			
			\item[]
			\FOR {$i$ = $1$:$Length(x_{sad_{new}})$}
			\STATE $possible_{unique\_min}(i)$ = Descend$(x_{sad_{new}}(i))$
			\item[]
			
			\IF {$possible_{unique\_min}(i)$ is unique}
			\STATE$x_{min_{temp}}$.append$(possible_{unique\_min}(i))$
			\ELSE
			\STATE continue
			
			\ENDIF
			\item[]	 
			\ENDFOR 
			\item[]
			
			\STATE $x_{min_{new}}$.append$(x_{min_{temp}})$  \%List of unexplored minima
			\STATE $x_{min_{unique}}$.append$(x_{min_{temp}})$ \%Full list of unique minima
			\STATE $x_{sad_{unique}}$.append$(x_{sad_{temp}})$ \%Full list of unique saddles
			\ENDWHILE

		\end{algorithmic}
		
	\end{algorithm}
	
	
	\subsection{Parallelizability}
	Two of the major benefit of this algorithm are its high parallelizability and its high degree of flexibility in both execution and application. Each minimum can be explored independently from the others, the strings all act independently from one another, and all of the images along each string act independently from each other during the gradient descent step. Thus, the while loop in Line 5 of Algorithm \ref{SSAlg} can be run as a parallel process, the for loops in lines 8 and 17 can also be run as parallel processes, and the string evolution in the \verb|Climb| and \verb|Descend| functions can have each image evolve in parallel. It is difficult to parallelize all of these process at the same time and is disallowed in many programming languages. So a choice of where to parallelize must be made. 
	
	Parallelizing over the while loop requires significant inter worker communication as it requires the list of potential new saddles and minima to be transferred at the end of each iteration. However, parallelizing here is beneficial for investigating smaller systems where the minima/saddle lists are small in memory or  for a system that is very large or has a complicated potential function such that the time spent finding the minima and saddles is relatively long compared to the time spent transferring data between workers. The two for loops are easier to parallelize and parallelizing here would be a natural choice for systems with simple potential functions where the time spent climbing and descending strings is relatively quick. These options in the implementation of the HSM allows it to be adapted to efficiently explore any system. 

	\section{Example System}	
	\label{Test System}
	
\begin{figure}[]
		\centering
		\includegraphics[width=0.9\linewidth]{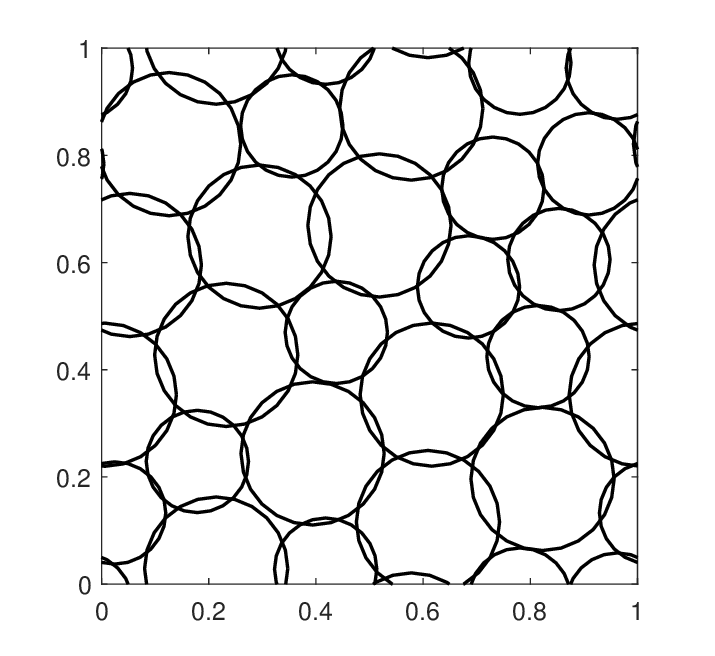}
		\caption{A sample energy minimizing configuration of the 24 Sphere System.}
		\label{fig:24SphereSystem}
	\end{figure}

	As a test of the HSM, we have applied it to a simple example system. This test system we analyze is made up of 24 bi-disperse soft spheres in a 2D periodic domain of unit length. The particles have a radii ratio of $1.4$ to prevent crystallization with half having a radius $R_L = 0.1336$ and half having a radius $R_S = 0.0954$. This arrangement of particles is shown in Figure \ref{fig:24SphereSystem}. 
	
	The periodic boundaries allow us to simulate a larger system with comparatively few particles. However, the periodicity introduces two symmetries to the system which are undesirable for the analysis we wish to perform; critical points become 2-dimensional sheets rather than points. These symmetries correspond to concerted shifts of all particles in the $x$ or $y$ directions. To break this symmetry, we fix one of the particles so it is not allowed to move even when other particles exert a force on it. This removes the symmetries and reduces the dimensionality of the system to $46$.
	
	The potential energy of the collection of these soft spheres is simply the sum of the pairwise spring potentials between the $N$ particles, given by:
	
	\begin{equation}
		\label{Potential Energy}
		V = \sum_{i = 1}^{N} \sum_{j = i}^{N} \boldsymbol{\kappa}_{i,j} \left (1 - \frac{|\textbf{x}_{i} - \textbf{x}_{j}|}{(R_{i}+R_{j})} \right )^2 J_{ij}.
	\end{equation} 

	In Eq. \ref{Potential Energy}, $N$ is the number of soft spheres, $\boldsymbol{\kappa}_{i,j}$ is the stiffness tensor (analogous to the spring constant) between particles $i$ and $j$, $\textbf{x}_i$ is the position vector of the center of particle $i$, $R_i$ is the radius of particle $i$, and $J_{ij}$ is zero if particles $i$ and $j$ do not overlap and one if they do. By setting $\kappa=1$ we find the energy-minimizing configurations, like the one shown in Fig. 3, appear similar to photoelastic disk systems studied by experimentalist studying shearing particles \cite{Majmudar_2005} (a review of the use of photoelastic particles in studying granular materials is available in \cite{Abed_Zadeh_2019}
	
	The number of minima of such a system scales exponentially with $N$ \cite{Ashwin_2012} and most physical systems will have $N$ much larger than the $24$ we have used here. This system is large enough to prohibit an exact enumeration of minima/saddles yet not untractable as a test system for the HSM.
	
	\section{Applying the Method}
	\label{sec:NumSims}
	
	The HSM requires several parameters, summarized in Table \ref{table:ParamTable}, to be set to efficiently generate an accurate and useful network. To choose these parameters, we need information about the ``basic structure" of the PES. By ``basic structure" we mostly refer to the structure of the basins of attraction around the minima in the PES since the Hydra String Method finds minima and first order saddle points which lie along the ridges of the basins. Of chief interest is the distance between basins, the radial size of the basins, and the number of saddles connected to a minimum. Knowing this structure tells us how far away from minima to look for saddles, how accurately we need to resolve the basin, and how many strings we need to extend from a minimum to find the saddles. 
	
\begin{center}
	\begin{table}[t]
		\caption{Hydra String Method Parameters}
		\begin{tabular}{| m{1.35cm} | m{5.15cm}| m{0.8cm}|} 
		\hline
		Parameter & Description & Value \\
		\hline
		\hline
		$\nu$ & controls how quickly the climbing image climbs towards a saddle, must be greater than 1  & $2$ \\
		\hline
		$dt$ & the time step used in the string evolution ODE solver & $10^{-3}$\\
		\hline
		$tol_{diff}$ & the tolerance used to stop the string evolution ODE solver & $10^{-8}$ \\
		\hline
		$n_{step_{max}}$ & the maximum number of time steps allowed in the string evolution ODE solver &  $10^5$ \\
		\hline
		$dist_{ext}$ & the distance a string is initially extended away from a minimum or saddle & $0.16$\\
		\hline
		$n_{images}$ & the number of images along a given string & $10$ \\
		\hline
		$n_{strings}$ & the number of strings to send out from each minimum to find new saddles & $24$ \\
		\hline
		$tol_{sad/min}$ & the tolerance to call a saddle/minimum unique & $10^{-2}$\\
		\hline
		
		\end{tabular}
		\label{table:ParamTable}
	\end{table}
\end{center}
	
	 Since the structure of the basins of an arbitrary function are generally not known a priori, we propose three numerical experiments as a way to deduce this basic structure. Each numerical experiment we propose addresses one of the characteristics of the basins we are interested in. We will make extensive use of these experiments to determine values for the various parameters used in the Hydra String Method which will be discussed in Section \ref{Parameters}.
	
	In our experience, determining this basic structure requires an iterative approach. That is, the easiest way to pick values for the parameters of the Hydra String Method is to first run the Hydra String Method. This can be done with very conservative choices for the various parameters which can then be used to refine those initial choices. This process can be repeated several times  to more fully understand the structure.
	
	In this section, we will show the results of these numerical experiments when run on the PES that arises from the system described in Section \ref{Test System}, a collection of 24 bi-disperse soft spheres that interact with pairwise linear spring forces contained in a unit box with periodic boundaries.	
	
	\subsection{Determining the Distance Between Basins}
		\label{PairwiseDist}
	
	When applying the Hydra String Method, the user must designate what the tolerance is to call minima(saddles) distinct. This arises because the climbing strings may approach the same saddle point or the hydra may curl back on itself and find the same minima over again. The strings may approach this minima(saddle) from a slightly different direction and due to the flatness of the PES at these critical points, they may approach slightly different values. One should appeal to the physics of the underlying system to help determine which minima(saddles) should be considered distinct. 
	
	To investigate this, we propose the following numerical experiment. Determine a collection of minima or saddle points. These can be found by either randomly sampling your PES and performing a gradient descent to find local minima. Or the Hydra String Method can be implemented with conservative values for the parameters required (small $dist_{ext}$, small $tol_{sad/min}$, large $n_{string}$, and large $n_{images}$) as mentioned above. These parameters will be discussed at length in the following section. With either approach, all that is necessary to perform the analysis is a collection of possibly degenerate critical points.
	
\begin{figure}[]
		\centering
		\includegraphics[width=\linewidth]{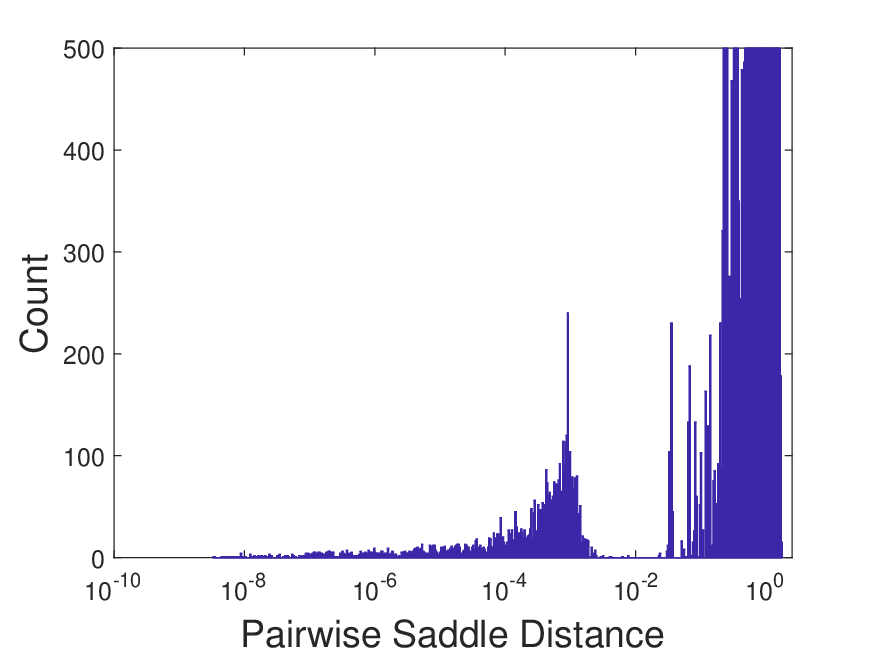}
		\caption{Histogram showing the pairwise distances between saddle points in a PES that arises from the system described in Section \ref{Test System}. The population between -1 and 0 are the clearly unique saddles and the population smaller than -3 are the degenerate saddles.}
		\label{fig:PairwiseDist}
	\end{figure}

	Now calculate the pairwise distance between all of the minima(saddles) and plot them on a histogram (as in \ref{fig:PairwiseDist}). Most of the pairwise distances should be quite large and indicate unarguably distinct critical points. A second population may arise near the stopping tolerance from the climbing string method ODE solver. Other populations may appear between these two pairwise distance groupings. By exploiting the physics that gives rise to the PES the user can set a minimum distance between basins(saddles) to call them unique.
	
	\subsection{Determining the Radial Size of a Basin}
		\label{BasinSize}
	
	Knowing the characteristic radial size of a basin allows the user to pick a distance to initially extend the climbing string. It also gives the user a general sense of the smallest features of interest of the PES. This knowledge can also be used to pick the number of images along each climbing string. These images are what ultimately detect these small scale features of the PES. 
	
\begin{figure}[]
			\centering
			\includegraphics[width=\linewidth]{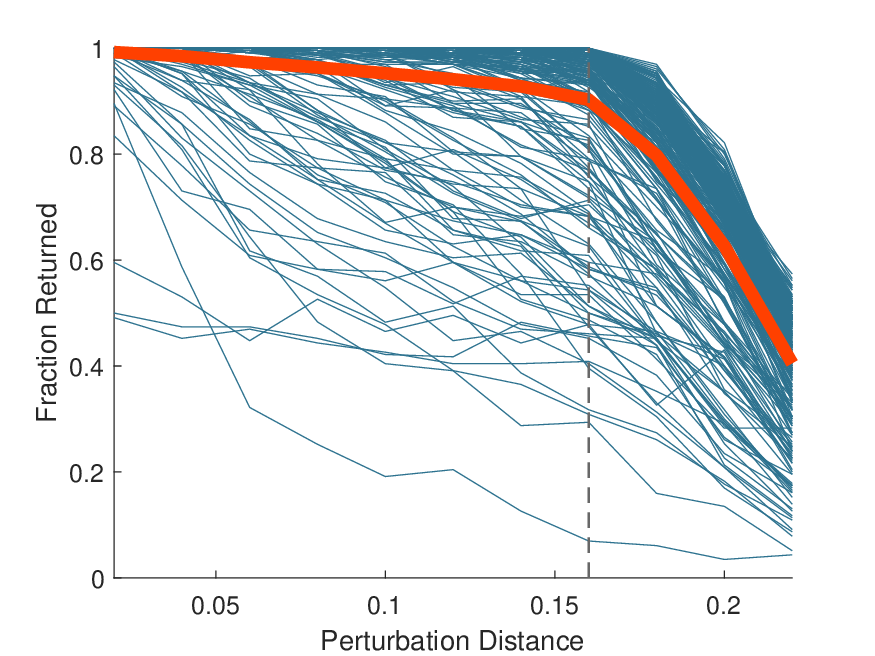}
			\caption{The results of a radial basin size analysis for the PES that arises from the system described in Section \ref{Test System}. Each thin blue-gray line represents a different minimum that was analyzed for its radial size. The thick red line is the average fraction returned at each perturbation distance across all basins. This analysis indicates most basins have a radius of approximately $0.16$, denoted by the dashed line, because beyond this perturbation distance, a rapidly increasing fraction of perturbed points fail to return to the original minimum. }
			\label{fig:RadialBasinAnal}
		\end{figure}

	To determine this characteristic radial size of the basins, we propose the following numerical experiment. Determine a representative collection of minima of the PES, as in Section \ref{PairwiseDist}. Perturb these minima a small distance from the minimal state in many different directions. Then take these perturbed points, gradient descend them, and determine if these perturbed points return to their original minimal state. Track how many perturbed points return to original minimum and repeat this for many different perturbation distances. This should be repeated for many different minima as the basins may not all be identical. These results can be plotted as they are in Fig. \ref{fig:RadialBasinAnal}.
	
	This sort of analysis can tell the user the size of the smallest features of interest in the PES. It can also reveal interesting features about the structure of the basins. This may be of interest in of itself, but this analysis will be necessary to implement the Hydra String Method because we need these results to determine both $dist_{ext}$ and $n_{images}$ which will be discussed in the next section.

	\subsection{Determining the Number of Connected Saddles}
		\label{NumSadAnal}	
	
	The Hydra String Method calls for extending some number of climbing strings from each minimum to find new saddles. To choose the number of strings to extend out, it is useful to know how many unique saddles are connected to a given minimum. That is, how many saddle points lie on the ridge of the basin of attraction for any given minimum. 
	
\begin{figure}[]
		\centering
		\includegraphics[width=\linewidth]{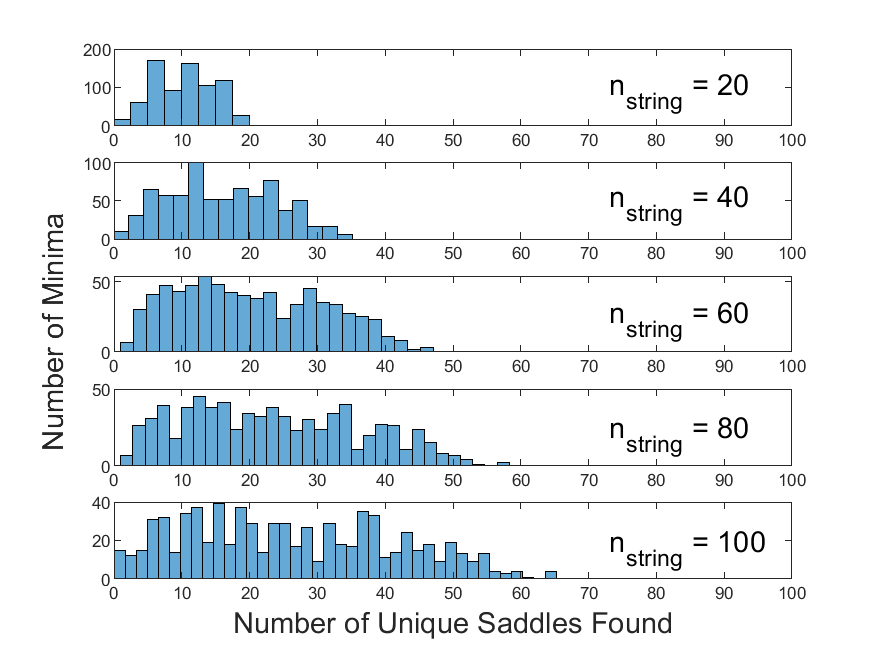}
		\caption{A series of histograms showing the number of minima from which a search was started and how many unique saddles were found connected to the minimum with a given number of strings extended.}
		\label{fig:SadHists}
	\end{figure}

	To determine this, we propose the following numerical experiment. Again, find a collection of minima to study. From each minimum extend out one string and note the saddle it converges to, then send a second, a third, etc. Note how many unique saddles are found for the number of extended strings. In practice, this can be accomplished by sending out a large number of strings all at once and determining how many converged to unique saddles after the fact. This should be done for a large number of minima because, as before, each basin need not be the same as the others. Plotting these data as a series of histograms  for various numbers of strings extended  as shown in Fig. \ref{fig:SadHists} with the number of unique saddles found on the x-axis and the number of minima with that many unique saddles found on the y-axis will give the user an idea of how many unique saddles the basins in the PES have.
	
	\section{Parameters Involved in the Hydra String Algorithm and How to Choose them\label{sec:Results}}
	
	Equipped with the numerical experiments from section \ref{sec:NumSims}, we will look at each parameter tabulated in Table \ref{table:ParamTable} and describe how to choose a value for it. Specifically, we will discuss how to choose these parameters for the system described in Section \ref{Test System}. We find it convenient to group them into three categories of parameters: those used to find a saddle, those used to ensure the saddle is in the original basin, and those used to generate a network of saddles and minima. We discuss each group in turn.

\subsection{Parameters Used to Find a Saddle}
\label{Parameters}

	These are the parameters needed to find an individual saddle, i.e. those needed to implement the climbing string method. They relate to numerically integrating the ordinary differential equation (ODE) describing the gradient descent of the images, and don't require specially designed tests to  find these parameters.

	\subsubsection{$\nu$}
	 The value of $\nu$ controls how quickly the climbing image climbs and must be greater than $1$, but not too large so as to cause the ODE solver to become unstable preventing the climbing image from converging. Too small a value will cause a slower convergence. We use the typical value of $2$ since this reflects the descent direction into the ascent direction, and the values of the other parameters can be tailored around this choice.
	
	\subsubsection{$dt$}
	The size of a time step of the climbing string and gradient descent ODE solver should be chosen by considering the normal accuracy, convergence, and stability issues: not too small to have an excessive time to convergence and not too large to become unstable. The value of $dt$ should also be chosen such that the maximum movement of an image along the string during a gradient descent is small enough to not jump over any important features of the energy landscape. We used $dt = 10^{-3}$.
	
	\subsubsection{$tol_{diff}$} \label{tol_diff}
	The criterion to stop the evolution of the climbing or descending string is set by the difference in euclidean distance of an image along the string between two consecutive time steps. That is, how much an image moves between two consecutive time steps (including both ODE movement and interpolation of the string images). This parameter should be smaller than the smallest feature of the PES one wishes to resolve but also large enough that the string converges in a reasonable amount of time. The results of the experiment in Section \ref{BasinSize} can help determine the sizes of these smallest features. We found that $~10^{-8}$ was a good balance between these conflicting goals.
	
	\subsubsection{$n_{step_{max}}$}	
	The maximum number of time steps allowed in the string evolution ODE solver needs to be large enough to allow convergence for most strings sent out during the snaking string algorithm. However, too large of a value leads to unnecessary computational load from runaway strings that have gone awry for one reason or another. This parameter is directly affected by $dt$, $\nu$, and $dist_{ext}$. Thus, it should be chosen after suitable values of those parameters have been chosen. After finding the maximum number of time steps needed for a typical string evolution, via computational trials, a moderate margin should be added to this maximum to get a starting value of $n_{step_{max}}$. The competing goals of minimizing computation time and allowing enough strings to converge to obtain a representative sample of the saddles in a system must be balanced in choosing a value of this parameter. We used $10^5$ in our system.
	
	\subsection{Parameters to Ensure a Saddle is in the Original Basin}
	
	These two parameters ensure that the string does not leave the basin of attraction from the originating minimum in the climbing string method. These two values vary based on a granular system's parameters such as particle radii, number of particles, etc. and will vary for different sorts of PESs. It is at this point that the analysis described in Section \ref{BasinSize} should be performed. Looking at the results (Fig.~\ref{fig:RadialBasinAnal}) of this analysis, we need to decide what the characteristic radial size of these basins is. As shown by the thick red line in Figure.~\ref{fig:RadialBasinAnal}, we see that $0.16$ appears to be the average radius of these basins. This radius was chosen because beyond this perturbation distance, a rapidly increasing fraction of perturbed points fail to return to the original minimum. Additionally, at this distance, for no minimum do all perturbed points return to the original minimum.

	\subsubsection{$dist_{ext}$} \label{distext}
	The distance by which a string is initially extended from a minimum(saddle) to find more saddles(minima) needs to be small enough to remain in the basin but also large enough to leave the region where the gradient is close to $0$ to minimize the time spent evolving to a saddle or minimum. A good choice is to pick the extension distance as the characteristic radial size of these basins. In this case, $dist_{ext} = 0.16$. One could also extend the string further than this radius if $n_{images}$ is chosen appropriately. If there are enough images, the basin is resolved well enough to locate its edge. If a string is initially extended beyond the original basin, it will be truncated leaving enough images in the original basin so that the string can be re-interpolated and continue climbing.
	
	\subsubsection{$n_{images}$}
	
	The number of images, $n_{images}$, should be chosen large enough to properly approximate the MEPs. However, too many images will lead to an increase in the time spent running the algorithm. For larger systems, this can quickly become computationally prohibitive. We recommend choosing $n_{images}$ such that the string resolves the smallest features of interest in the landscape. Figure \ref{fig:RadialBasinAnal} shows that there are larger and smaller basins, so sending out a string the distance of the average basin size may miss these smaller basins. To prevent this, one should interpolate many images along the string to capture these smaller features. In this case, we chose $n_{images} = 10$ so that an image falls inside these smaller basins; we potentially ignore some of the very small basins. 
	
	In this case, most of the basins have approximately the same structure. That is, they are relatively flat and spherical up until about $0.16$. If instead there were many basins with different structures, one may need to have more images to accurately capture the complicated structure that appears on the ridges of these basins. Our basins are quite uniform and so we can choose a small number of images without missing many small features. We may not accurately resolve the MEP but we are confident we find the minima and saddles of interest to our analysis. If one desires a more accurate MEP, they should interpolate many more images to resolve it more finely. 
	
	\paragraph*{The Interplay between $dist_{ext}$ and $n_{images}$}
	Since the energy of the climbing string at each image is kept monotonic by cutting the string if it ever fails to maintain monotonicity, the $dist_{ext}$ can be chosen to be much larger than the radial size of the smallest basin as long as the string resolves the smallest basins of interest. For example, the $dist_{ext}$ can be chosen to be $0.2$ instead of $0.16$. We can than interpolate more images, say $n_{images} = 15$, and if the string leaves a smaller basin, the string should detect that it has left the basin, because the energy fails to be monotonic, and cut itself. Overestimating the $dist_{ext}$ allows the strings to quickly leave the region with a small gradient and having a large $n_{images}$ prevents the over-extended string from leaving the basin.

	\subsection{Parameters to Make a Network of Saddles}
	
	The Hydra String Method generates a network where saddles and minima are nodes and the minimum energy paths between them are the edges. To generate this network, we need to decide which minima and saddles are unique from one another (as discussed in Sec.~\ref{PairwiseDist}) and we need to find the saddles between these minima.
	
	\subsubsection{$tol_{sad / min}$} \label{tol_sad}
	
	Often, the Hydra String Algorithm finds minima and saddles that are very near to each other in euclidean space with nearly exactly the same energies. The question arises, are these differences from numerical tolerances or are they in fact distinct points? To answer this we proposed the numerical experiment in Section \ref{PairwiseDist} to determine a numerical tolerance to call these granular configurations degenerate. 
	
	The results of such a numerical simulation are shown in Fig. \ref{fig:PairwiseDist}. Due to the number of saddles included, we zoomed in to clearly see the important features. The region with pairwise distances between $~10^{-1.5}$ and $~10^{0}$ are the obviously distinct saddles which can be seen by plotting the two configurations for a given pair in this region. The region between $~10^{-2.5}$ and $~10^{-8}$ are the obviously degenerate saddles. Again this can be seen by plotting both configurations for any given pair in this region and noting they overlap almost perfectly. From this, we chose $tol_{sad / min}$ to be $10^{-2}$. All configurations closer than $10^{-2}$ in euclidean distance would be considered degenerate.
	
	This histogram is likely to appear wildly differently for different PESs. In the case of a granular material, these pairwise differences will change in magnitude based on the radii of the particles and the number of particles themselves, and thus the dimension of the system. The distance function begins to be less useful at higher dimensions and so different means to differentiate saddles and minima may be necessary for sufficiently large systems. One possible substitute for the case of granular materials is to look at the contact matrices of the constituent particles and defining configurations with different contact matrices to be different. 
	
	It is important to note how variable this histogram may appear because this numerical experiment requires the user to apply some knowledge about the PES and the physics of the system that gives rise to that PES to interpret this plot and choose a value for $tol_{sad / min}$. We were able to apply physical intuition about the grains, whether the configurations were imperceptibly different to the human eye and the contact matrices, to help us choose the value of this parameter. Systems described by different physical laws will require different considerations. 
	
	\subsubsection{$n_{strings}$}
	The number of strings to be sent out from each minimum to find new saddles should be large enough to find a statistically representative sample of the saddles connected to a minimum but small enough that not many of the strings converge to the same saddle. The analysis described in Section \ref{NumSadAnal} should be performed to help choose the value of this parameter. The number of cores on the system should also be considered when choosing a value of $n_{strings}$, due to the parallel nature of the snaking algorithm, a multiple of the number of cores may make the execution more efficient.
	
\begin{figure}[]
		\centering
		\includegraphics[width=\linewidth]{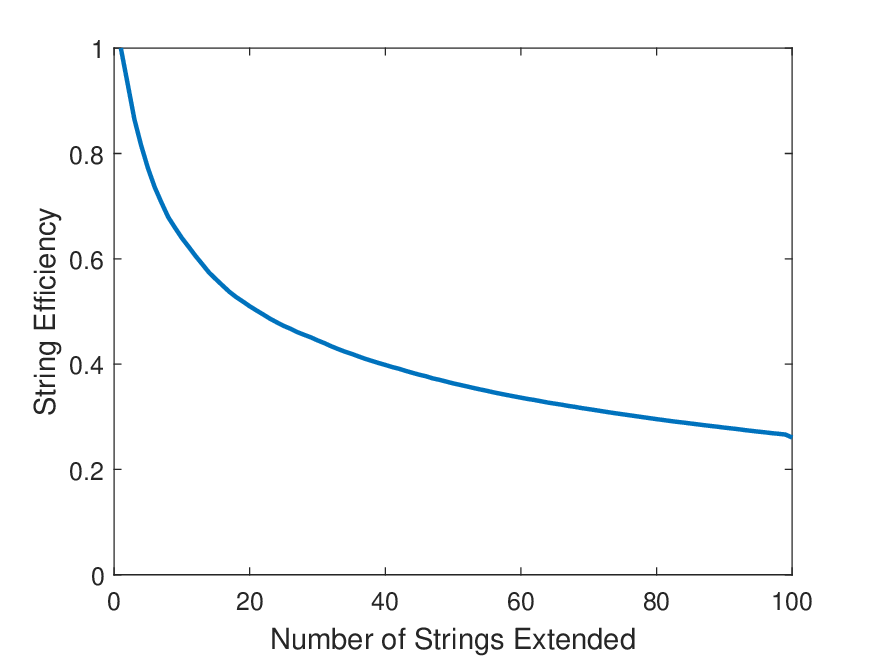}
		\caption{This plot shows the efficiency of the strings sent to find unique saddles around a basin. The y-axis is what fraction of strings converged to new saddles and the x-axis is how many strings were extended. This is an average over 750 minima.}
		\label{fig:MeanSadEff}
	\end{figure}

	Following the method outlined in Sec. \ref{NumSadAnal}, five histograms are shown in Figure \ref{fig:SadHists} generated for $20, 40, 60, 80$ and $100$ strings. The histograms might immediately indicate how many unique saddles exist on the ridge of a basin. However, we have found that as more strings are extended, there are generally more saddles to be found. We instead find it useful to look at the efficiency of these strings which we define as the fraction of strings that find unique saddles. This efficiency is plotted in Figure \ref{fig:MeanSadEff}. Choosing an efficiency of $50\%$ (half the maximum) would dictate $n_s = 20$. Our computing hardware has nodes with $12$ or $24$ cores, so we chose $n_{strings} = 24$ to take advantage of the parallelizability of the Hydra String Method.

\begin{figure}[]
		\centering
		\includegraphics[width=\linewidth]{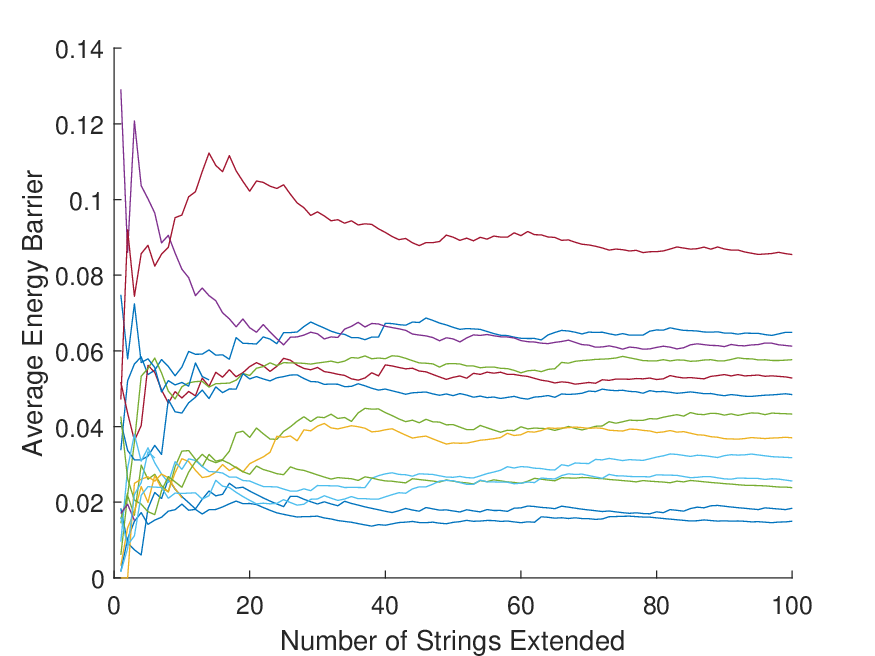}
		\caption{This plot shows the average energy barrier between a minimum and the saddles found in the number of unique saddles analysis as a function of the number of extended strings. Each line represents a different minimum explored. Only approximately 20 lines of the 750 are shown for clarity.}
		\label{fig:AvgErgBar}
	\end{figure}

	One may also apply an analysis that appeals to the underlying physics of the system to determine the number of extended strings. For example, in our future work with the Hydra String Method, we wish to find various transition paths the system might follow as it moves between minima. We predict the probability that the system moves through these saddles is related to the energy barrier between the minimum and the saddle. So, we may want to generate a plot of these energy barriers as a function of the number of strings extended as shown in Figure \ref{fig:AvgErgBar}. From this figure, we can see that the average energy barrier found levels off around $30$ or so extended strings. As before multiples of 12 are convenient so we might choose $n_{strings} = 36$ in this case. 

	In this system, the histograms from Figure \ref{fig:SadHists} each appear largely uniformly distributed. However, we have also analyzed systems with bi-modal or heavily tailed distributions. So, one should look at these underlying histograms before immediately creating the efficiency plot to find a value for $n_{strings}$ as one may need a different measure of the average than a simple mean to effectively create this efficiency plot. 
	
	One must also consider what is more important in their analysis of their PES. Is it more important to find every connection between all minima of the PES? Or does one favor exploring more of the PES at the cost of missing a few saddles and minima? In our case, we were not concerned with finding every hard to find saddle but we do want to find the many possibly interconnected saddles between minima. So, we chose an intermediate goal, finding many but not all minima and the connecting saddles. 
	
	\FloatBarrier		
	
	\section{Discussion\label{sec:Discussion}}
	
	Choosing the various parameters in the Hydra String Method allows it to be adapted to different goals such as short range accuracy or long range exploration.  Thus one can choose if it is more important to painstakingly map out every minimum and saddle point in your network? Or is it more important to find many minima and saddle points far from the initial point from which the Hydra String Method begins? These conflicting goals appear several times in the Hydra String Method. Does one choose a short extension distance to ensure no nearby saddles are missed? Or does one pick a large value to quickly find further away saddles? Does one extend a large number of strings to find as many connected saddles as possible? Or a few to find the most common saddle points and move further into the PES? And so on.
	
	In our case, we are analyzing the PES of a collection of soft spheres to map out various transition pathways the system can undergo. That is, starting in a stable configuration, what are the nearby stable configurations and what unstable configuration, saddle point, do they pass through as the system moves between these stable configurations? In this application, we favor exploring more of the PES instead of finding every possible minimum or saddle point. Many of those configurations may be unlikely for the granular system to reach and are therefore unimportant to our future analyses. 
	
	However, for other systems the Hydra String Method might be applied to, it may be more important to locate every minimum or saddle. For example, when studying a chemical system, one might be interested in determining various by-products or possibly dangerous intermediate products of a chemical reaction. In that case, it might be more important to find every possible minimum or saddle point.
	
	\section{Conclusion\label{sec:Conclusion}}

	In this paper we presented the Hydra String Method, a novel computational method to autonomously and efficiently map the minima and first order saddle points of a PES. In doing so, we presented a systemic approach to tailor the various system specific parameters of the HSM to arbitrary systems and demonstrated this approach on an example soft-sphere granular system. The results of the above numerical experiments may be very different for other systems with different potential functions, such as a pairwise Lennard-Jones potential between particles which is commonly used in bubbles \cite{Zhang_2017} and as a model for molecules/atoms in chemistry\cite{Chill_2014}. This broad applicability is one of the major assets of this method.
	
	In our case, we intend to apply this method to granular systems to map out the various stable configurations and determine the MEPs that connect them. We believe that, these MEPs approximate the transition paths the system undergoes when sheared slowly and with sufficient damping.  We hypothesize that the transitions with the lowest energy barriers will be the most likely transitions the system will undergo when slowly sheared with sufficient damping.
	
	Since this method utilizes the String Method, it inherits all of the benefits that the String Method has over other saddle finding schemes. Of note are the modular re-interpolation of images along the string and the improved stability. As previously discussed, the interpolation of the images along the string can be easily changed from a linear interpolation to a cubic spline or any other method. If desired, the images can be ``bunched up" around areas where the landscape may have more intricate structure or ``thinned out" in relatively barren regions.
		
	Finally, the advantage this method gains from its highly parallelizable nature and autonomous execution can hardly be overstated. The method efficiently realizes computational speed ups with growing numbers of parallel cores. The time spent climbing and descending strings is much greater than the time spent with overhead memory transference and using a list of unexplored minima to direct the workers mitigates wasted computation time spent searching already explored regions of the PES. The upfront time spent in picking suitable parameters for the method is quickly recouped in the autonomous execution of the search. With some intuitive precautions such as a maximum number of time steps allowed on the climbing/descending functions (to prevent strings from becoming stuck) and a maximum allowed energy along a string (to prevent run away strings) the Hydra will happily branch out and explore any energy surface.
	
	\section*{Acknowledgements} 
	
	The authors thank Eric Vanden-Eijnden for discussions on the implementation of the string method as well as Karen Daniels, Ryan Kozlowski and Jack Featherstone for their feedback on an earlier draft of this manuscript.  
	
	\section*{Declarations}

	\textbf{Funding} This work was supported by the National Science Foundation DMS-1816394.

	\section*{Compliance with Ethical Standards}

	\noindent\textbf{Conflict of interest} The authors declare that they have no conflict of interest\\

	\noindent\textbf{Availability of data and material} All processed data is presented in the manuscript figures.\\

	\noindent\textbf{Code availability}\\
	\href{https://github.com/knewhall/Hydra\_String}{https://github.com/knewhall/Hydra\_String}
	

\end{document}